\documentclass[12pt]{article}
\usepackage[dvips]{color}
\usepackage{epsfig}
\usepackage{amsmath}
\usepackage{graphicx}

\begin{document}
\title{\bf P - V Criticality of Logarithmic Corrected Dyonic Charged AdS Black Hole}
\author{J. Sadeghi$^{a}$\thanks{Email: pouriya@ipm.ir}\hspace{1mm} B. Pourhassan$^{b}$\thanks{Email: b.pourhassan@du.ac.ir}\hspace{1mm}, M. Rostami $^{c}$\thanks{Email: m.rostami@iauamol.ac.ir}\\
$^{a}${\small {\em  Department of Physics, University of Mazandaran, Babolsar, Iran}}\\
$^{b}${\small {\em  School of Physics, Damghan University, Damghan, Iran}}\\
$^{c}${\small {\em  Department of Physics, Tehran Northem Branch, Islamic Azad University, Tehran, Iran}}}

\maketitle

\begin{abstract}
\noindent

In this paper, we consider dyonic charged AdS black hole which is holographic dual of a van der Waals fluid. We use logarithmic corrected entropy and study thermodynamics of the black hole and show that holographic picture is still valid. Critical behaviors and stability also discussed. Logarithmic corrections arises due to thermal fluctuations which are important when size of black hole be small. So, thermal fluctuations interpreted as quantum effect. It means that we can see quantum effect of a black hole which is a gravitational system.\\\\

{\bf Pacs Number:}  04.60.-m 04.70.-s 05.70.-a

\end{abstract}

\section{Introduction}
We know that the black hole entropy is proportional to the horizon aria $A$, known as entropy-area law of the black hole \cite{1}. On the other hand, thermal fluctuations will happen for any thermodynamical system like black holes. Specially, thermal fluctuations are more important for the small objects, because one can see that thermal fluctuations arise due to quantum fluctuations in the geometry of space-time. So, one can neglect such fluctuations for the large black holes.
However, when the size of black hole decreased due to Hawking radiation, the effect of quantum fluctuations on the geometry of the black
holes will be important. Thermal fluctuations modify thermodynamic of system and they appear in the black hole entropy as logarithmic term, So logarithmic corrected entropy may be written as \cite{2, 3},
\begin{equation}\label{s1}
S = \frac{A}{4}-\frac{\alpha}{2}\ln{(\frac{AT^{2}}{4})},
\end{equation}
where $\alpha$ is a constant usually fixed as unity. We can track effect of thermal fluctuations using $\alpha$ and obtain ordinary thermodynamics by setting $\alpha=0$. Recently effect of thermal fluctuations considered for several kinds of black objects like charged AdS black hole \cite{4}, black Saturn \cite{5, 6} and modified Hayward black hole \cite{7}. Moreover, in the Ref. \cite{sarkar} black
hole entropy corrections due to thermal fluctuations in the black hole
extensive parameters for various AdS black holes have been studied for both canonical and grand canonical
ensembles. Some asymptotic AdS black holes like BTZ, D=4 KN-AdS and R-charged black holes in various dimensions considered in that paper and found an universality in the logarithmic corrections to
charged AdS black holes entropy in various dimensions, which are expressed in terms of the black hole response
coefficients via fluctuation moments.\\
In this paper, we would like to consider an interesting kind of black hole which is called dyonic charged AdS black hole in four dimensions which has both electric and magnetic charges \cite{8, 9}. In the Ref. \cite{mann} using critical point investigation it is found that charged AdS black holes are corresponding to van der Waals fluid. Also, it has been argued that dyonic charged AdS black hole may be holographic dual of van der Waals
fluid with chemical potential \cite{10}. It is also found that van der Waals fluid is holographic dual of RN AdS black hole \cite{wu}. So, by using holographic principles one can study dyonic charged AdS black hole via a van der Waals. We use this motivation to study logarithmic corrected thermodynamics of dyonic charged AdS black hole to see such quantum gravity effect.\\
The paper is organized as follows. In the next section we give brief review of dyonic charged AdS black hole and recall necessary relations which need to study its thermodynamics under thermal fluctuation effect. In section 3 we obtain several thermodynamics relations and see the effect of logarithmic correction. In section 4 we use holographic picture which is a van der Waals fluid and study about critical behavior. In section 5 we discuss about stability of black hole, and finally in section 6 we give conclusion.

\section{Dyonic charged AdS black hole}
Dyonic charged AdS black hole is solution of Einstein-Maxwell
theory with negative cosmological constant in four dimensions which described by the following action \cite{9, 10},
\begin{equation}\label{s2}
I = \frac{1}{16\pi} \int d^4 x \sqrt{-g} \left ( -R - \frac{6}{l^2}
+ \frac{1}{4} F_{\mu\nu} F^{\mu\nu}\right ),
\end{equation}
where $R $ is the Ricci scalar, $F_{\mu\nu}$ is the strength of electromagnetic field and $l$ is the curvature radius of AdS space, also $G=1$ assumed. Solving Einstein equations yields to the following metric,
\begin{equation}\label{s3}
ds^2 = -f(r) dt^2 + \frac{dr^2}{f(r)} + r^2 d\theta^{2} + r^{2}\sin^{2} {\theta } d\phi^{2},
\end{equation}
where,
\begin{equation}\label{s4}
f(r) = 1 + \frac{r^2}{l^2} - \frac{2M}{r} + \frac{q^2_{e} +q^2_{m}}{r^2}.
\end{equation}
where $q_e$, $q_{m}$ and $M$ are electric charge, magnetic charge and mass of the
black hole respectively. Electric and magnetic potential  $\Phi$  defined by the following
equation,
\begin{eqnarray}\label{s5}
\Phi_{e} &=& \frac{q_{e}}{r_{+}},\nonumber\\
\Phi_{m} &=& \frac{q_{m}}{r_{+}},
\end{eqnarray}
where the black hole horizon $r_{+}$ is given by,
\begin{equation}\label{s6}
f(r) = 1 + \frac{r^2}{l^2} - \frac{2 M}{r} +
\frac{q^2_{e} +q^2_{m} }{r^2}=0,
\end{equation}
which has at least two real and positive roots $r_{\pm}$. So, one can write the mass of the black hole as follow,
\begin{equation}\label{s7}
M = \frac{r_+}{2} + \frac{r_{+}^{3}}{2 l^2} +  \frac{q^2_{e}
+q^2_{m} }{2 r_{+}}.
\end{equation}
$f(r)$ in terms of $r$, and black hole mass as a function of $r_{+}$ plotted by the Fig. \ref{fig1}. We will discuss about them in the next section when study the logarithmic corrected thermodynamics of the solution discussed above.\\
Before end of this section it is useful to note that 4 dimensional dyonic charged AdS black hole in asymptotically AdS space-time is conjectured to
be dual to a 3 dimensional CFT living on the boundary of the AdS space \cite{mann, 10}. The bulk gauge field is dual to a global $U(1)$ current operator $J_{\mu}$, while conserved global charge of CFT side given by,
\begin{equation}\label{Ap1}
J^{t}=\frac{q_{e}}{16\pi G},
\end{equation}
where $G$ is four dimensional Newtonian constant related to degree of the gauge group $N$ of the CFT using the holographic dictionary,
\begin{equation}\label{Ap2}
\frac{1}{4 G}=\frac{\sqrt{2}N^{\frac{3}{2}}}{6b^{2}},
\end{equation}
where $b^{2}$ related to the strength of the magnetic field $B b^{2}=q_{m}$.

\section{Logarithmic corrected thermodynamics}
\begin{figure}[h!]
 \begin{center}$
 \begin{array}{cccc}
\includegraphics[width=50 mm]{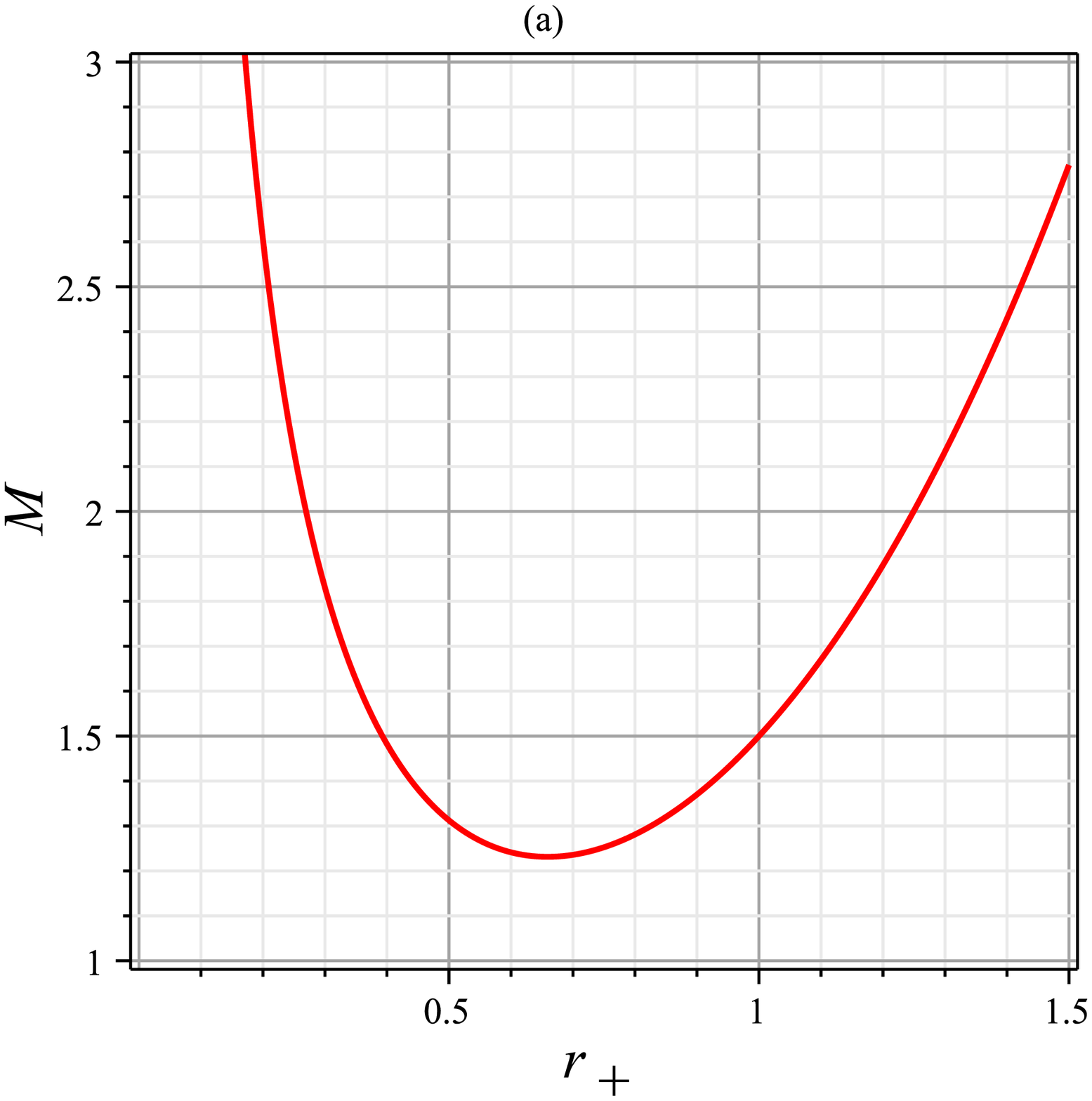}\includegraphics[width=50 mm]{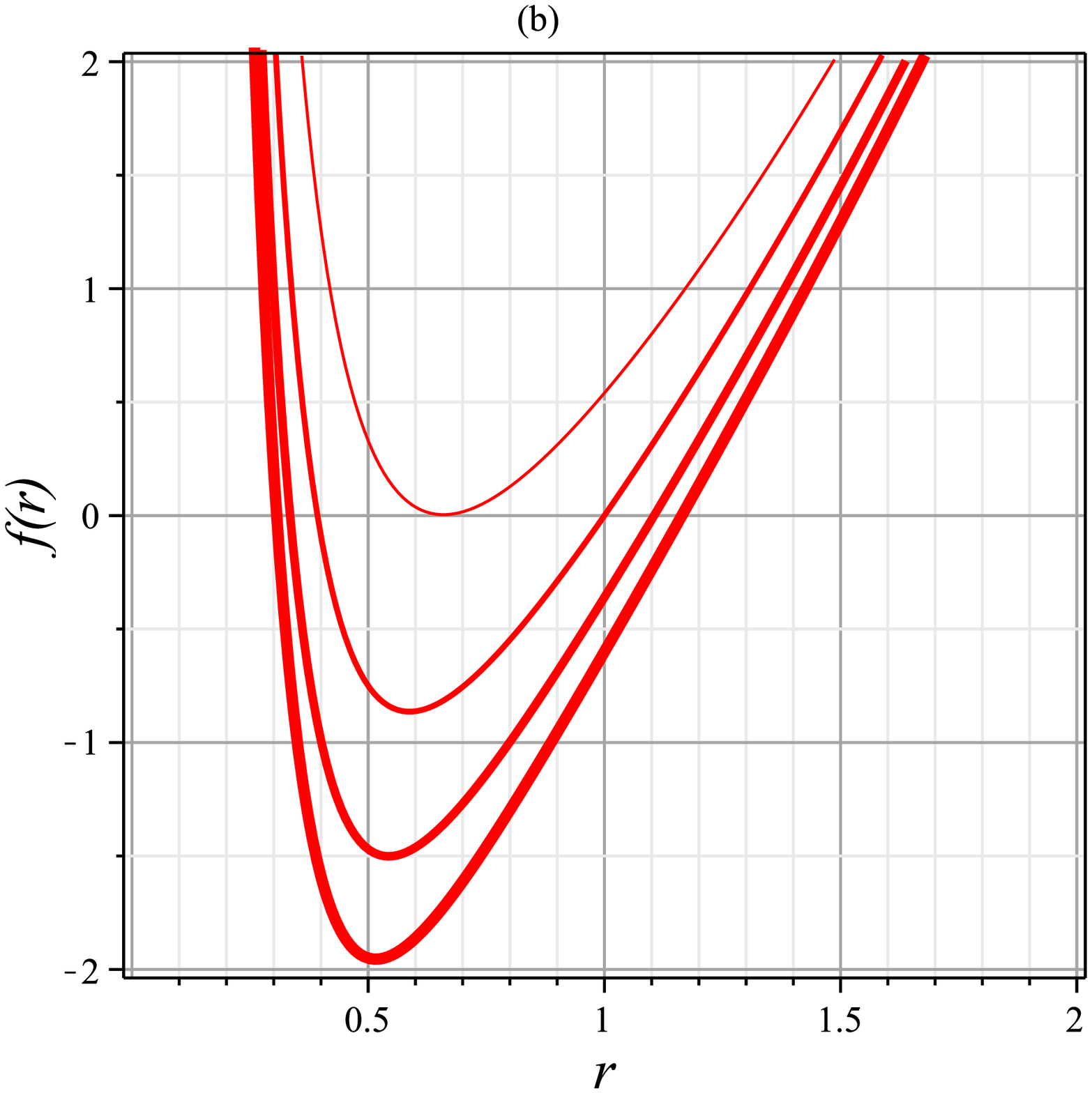}
 \end{array}$
 \end{center}
\caption{(a) Black hole mass in terms of event horizon. (b) Horizon radius with variation of black hole mass $M=1.23$ (thin), $M=1.5$, $M=1.68$, $M=1.8$ (thick). $l=q_{e}^{2}+q_{m}^{2}=1$.}
 \label{fig1}
\end{figure}

We know that a black hole may be considered as thermodynamic systems characterized by a
Hawking temperature $T$ and the corresponding logarithmic corrected entropy $S$ given by the equation (\ref{s1}), with $A=4\pi r_{+}^{2}$, hence we have,
\begin{equation}\label{s8}
S = \pi r_{+}^{2}-\frac{\alpha}{2}\ln{(\pi r_{+}^{2}T^{2})},
\end{equation}
where
\begin{equation}\label{s9}
T = \left(\frac{f^{\prime} (r)}{4 \pi }\right)_{r=r_{+}} = \frac{1}{4\pi}\left ( \frac{3
r_+}{l^2} + \frac{1}{r_{+}} -  \frac{q^2_{e} +q^2_{m}
}{r_{+}^3}\right),
\end{equation}
where we used the equation (\ref{s7}) in the last equality.\\
Using the entropy and temperature we can find the Helmholtz function,
\begin{equation}\label{s10}
F=-\int{SdT}=-\frac{r_{+}^{3}}{4l^{2}}+\frac{r_{+}}{4}+F(q)+F(\alpha),
\end{equation}
where
\begin{eqnarray}\label{s11}
F(q)&=&\frac{3(q_{e}^{2}+q_{m}^{2})}{4r_{+}},\nonumber\\
F(\alpha)&=&\frac{\alpha(q_{e}^{2}+q_{m}^{2})}{8\pi r_{+}^{3}}\left[\frac{4}{3}+\ln{\frac{8\pi l^{4}r_{+}^{4}}{(l^{2}(q_{e}^{2}+q_{m}^{2})-l^{2}r_{+}^{2}-3r_{+}^{4})^{2}}}\right]\nonumber\\
&-&\frac{3\alpha r_{+}}{8\pi l^{2}}\left[4+\ln{\frac{8\pi l^{4}r_{+}^{4}}{(l^{2}(q_{e}^{2}+q_{m}^{2})-l^{2}r_{+}^{2}-3r_{+}^{4})^{2}}}\right]\nonumber\\
&-&\frac{\alpha}{8\pi r_{+}}\left[\ln{\frac{8\pi l^{4}r_{+}^{4}}{(l^{2}(q_{e}^{2}+q_{m}^{2})-l^{2}r_{+}^{2}-3r_{+}^{4})^{2}}}\right].
\end{eqnarray}
Neglecting thermal fluctuation ($\alpha=0$) equation (\ref{s11}) vanishes. Also, the first term of r.h.s of the equation (\ref{s10}) is corresponding to ordinary AdS black hole with thermodynamics pressure related to the cosmological constant \cite{ray},
\begin{equation}\label{s12}
P(\alpha=q_{e}=q_{m}=0) = - \frac{\Lambda}{2l^2} = \frac{3}{16\pi}\frac{1}{l^2},
\end{equation}
and thermodynamic volume,
\begin{equation}\label{s13}
V =  \frac{4}{3}\pi r_{+}^{3}.
\end{equation}
Hence, in presence of electric and magnetic charge and thermal fluctuations, at least on of the above equations should be modified including charge and $\alpha$ dependent terms. We will assume thermodynamic volume given by the equation (\ref{s13}) and obtain modified pressure.\\
Let us now discuss about Helmholtz free energy given by the equation (\ref{s10}). We will fix parameters as $l=1$ and $q_{e}^{2}+q_{m}^{2}=1$. In that case from the Fig. \ref{fig1} (a) we can see that if $0.3\leq r_{+}\leq 1.2$ then $1.2<M<2$. So, from the Fig. \ref{fig1} (b) one can see extremal limit given by $M=1.23$, while $r_{+}=1$ is corresponding to $M=1.5$ or $M=1.68$ is corresponding to $r_{+}=1.1$.\\

\begin{figure}[h!]
 \begin{center}$
 \begin{array}{cccc}
\includegraphics[width=50 mm]{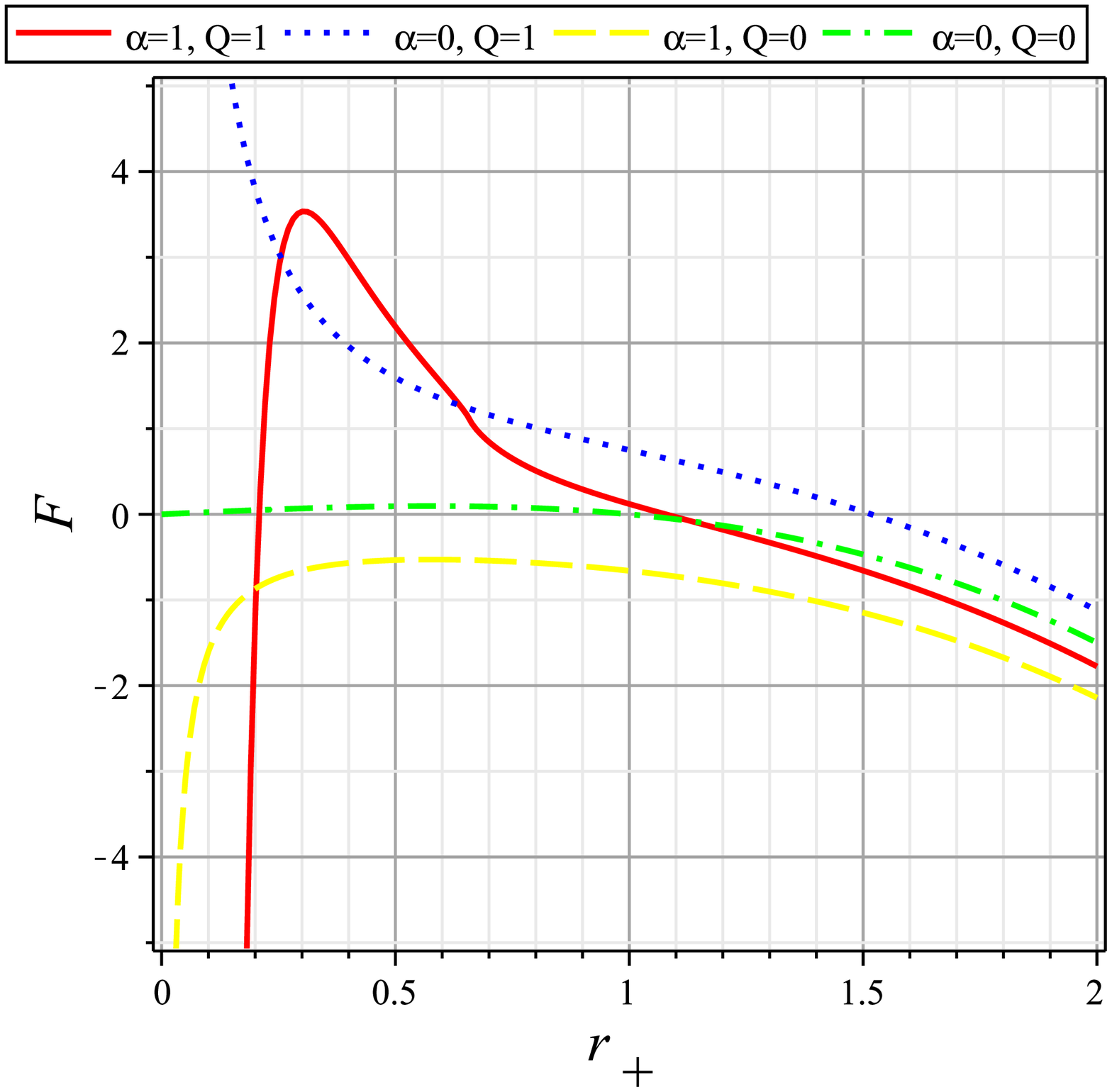}\includegraphics[width=50 mm]{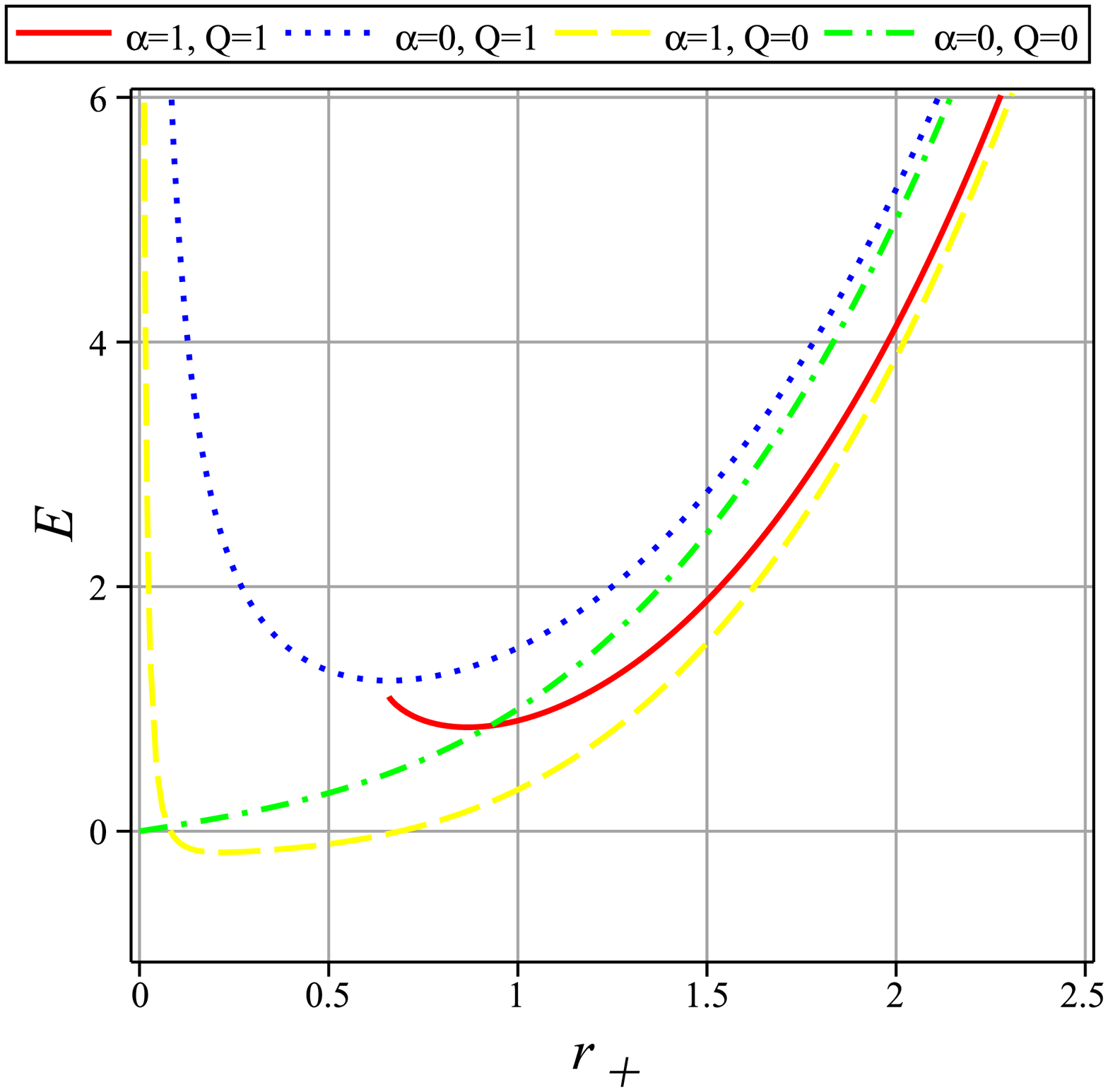}
 \end{array}$
 \end{center}
\caption{Left: Helmholtz free energy in terms of horizon radius with $l=1$. Right: Internal energy in terms of horizon radius with $l=1$. $Q\equiv q_{e}^{2}+q_{m}^{2}=1$, $\alpha=1$ (solid), $\alpha=0$ (dotted). $Q=0$, $\alpha=1$ (dashed), $\alpha=0$ (dash dotted).}
 \label{fig2}
\end{figure}

\begin{figure}[h!]
 \begin{center}$
 \begin{array}{cccc}
\includegraphics[width=50 mm]{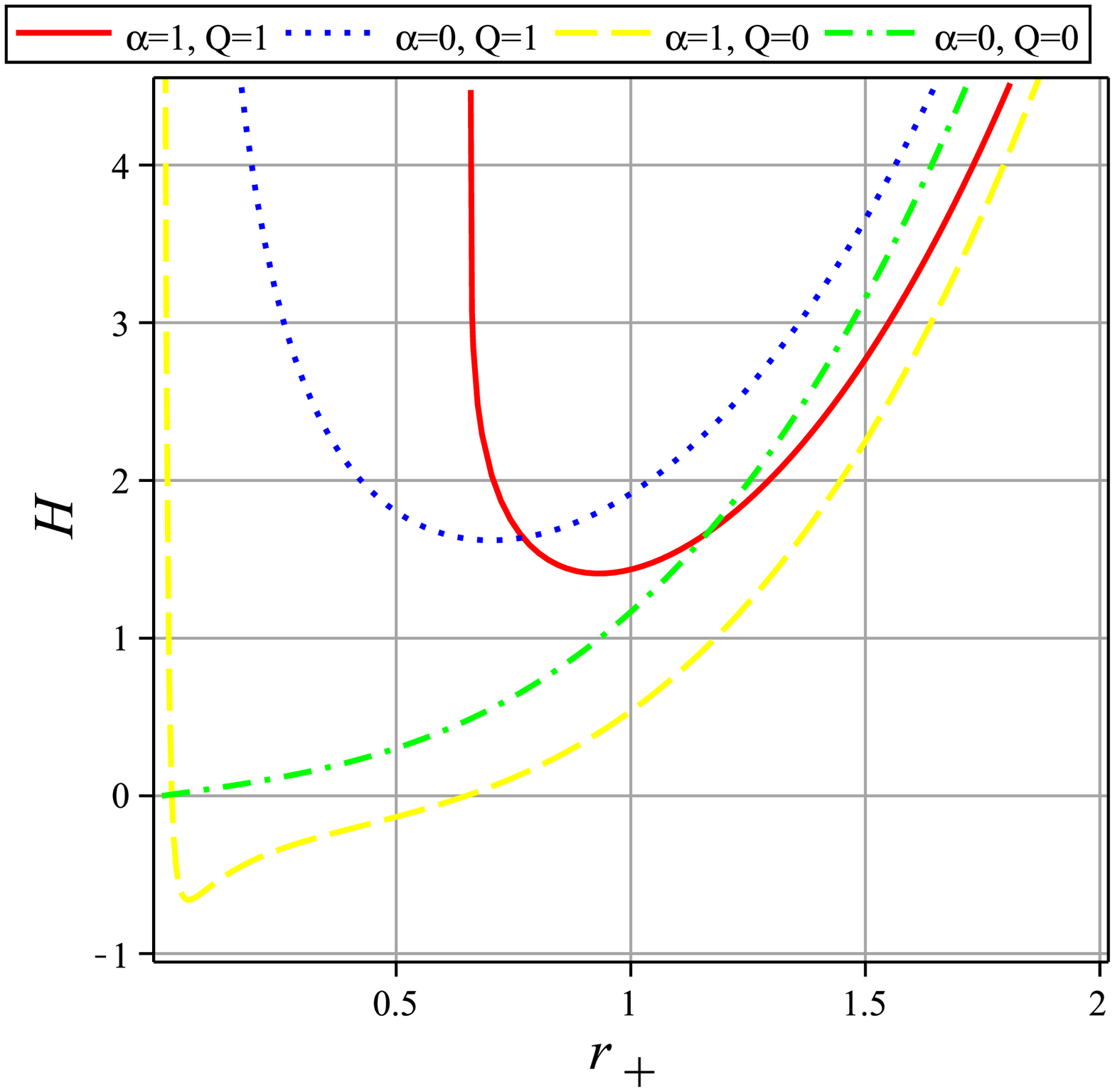}\includegraphics[width=50 mm]{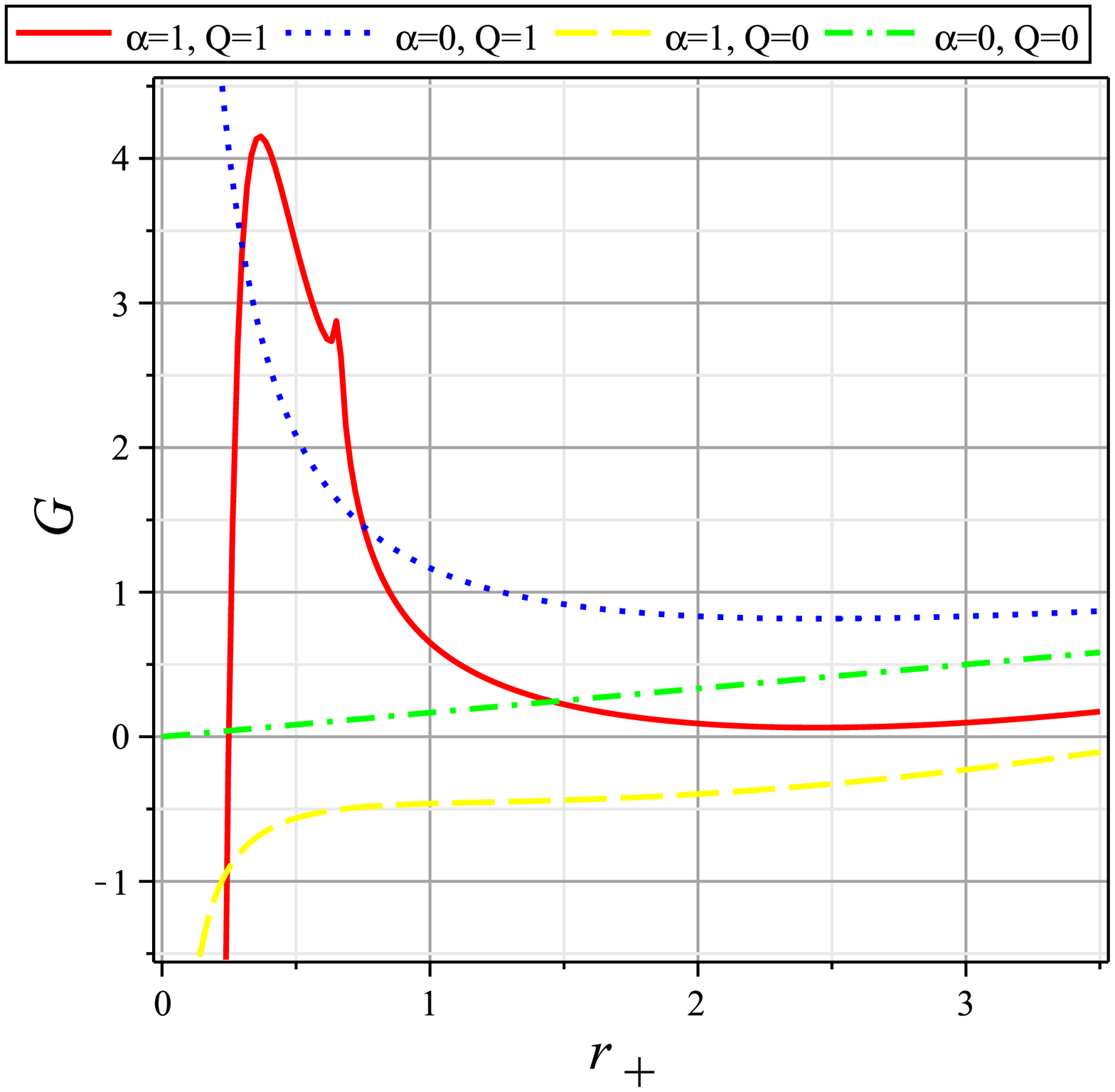}
 \end{array}$
 \end{center}
\caption{Left: Enthalpy in terms of horizon radius with $l=1$. Right: Gibbs free energy in terms of horizon radius with $l=1$. $Q\equiv q_{e}^{2}+q_{m}^{2}=1$, $\alpha=1$ (solid), $\alpha=0$ (dotted). $Q=0$, $\alpha=1$ (dashed), $\alpha=0$ (dash dotted).}
 \label{fig3}
\end{figure}

Then, from the Fig. \ref{fig2} we can see four different situations for the Helmholtz function and internal energy, all have similar behavior for the large black hole with $r_{+}\gg1$ where $F$ has negative large value and $E$ has positive large value. Effect of thermal fluctuations on the black hole with small size is obvious and important. Solid (red) and dashed (yellow) lines represent logarithmic corrected cases (charged and uncharged respectively) where Helmholtz function goes to negative infinity at $r_{+}\rightarrow 0$ limit, while for the case of $\alpha=0$ we have $F\rightarrow\infty$ (charged) and $F\rightarrow0$ (zero-charge) at $r_{+}\rightarrow 0$ limit. The solid red line may be corresponding to the black hole with mass $M=1.68$ where $F=0$ near the outer horizon ($r_{+}\approx1.1$) and $F=F_{max}$ near the inner horizon (see Fig. \ref{fig1} (b)). Outside of the black hole, free energy is negative for the mentioned black hole mass. We can see critical radius $r_{+c}\approx0.7$, before it internal energy vanishes (Fig. \ref{fig1} (b)) and Helmholtz energy have different behavior. As we will see later it relate to stability of black hole, ie. black hole is stable for $r_{+}>r_{+c}$. It means that the logarithmic effect is more important for small black hole.\\
In order to obtain internal energy we used well-known thermodynamics relation,
\begin{equation}\label{s14}
E=F+ST.
\end{equation}
Effect of thermal fluctuation on the internal energy is decreasing its value for the large black holes. On the other hand for the infinitesimal black hole ($r_{+}\rightarrow0$) internal energy will be infinite (see dashed yellow line of right plot of the Fig. \ref{fig2}). In the case of $\alpha=q_{e}^{2}+q_{m}^{2}=l=1$ the internal energy will be imaginary for $r_{+}<0.66$, which may be interpreted as critical value of horizon radius where black hole can exist. There is also a minimum for the internal energy corresponding to minimum mass shown in the Fig. \ref{fig1} (a).\\
As we told already, modified pressure due to thermal fluctuation can be obtained using the derivative of Helmholtz function with respect to the volume,
\begin{equation}\label{s15}
P=-\left(\frac{\partial F}{\partial V}\right)_{T}.
\end{equation}
By using the equations (\ref{s11}) and  (\ref{s13}) one can obtain pressure of dyonic charged AdS black holes as follow,
\begin{eqnarray}\label{s16}
P&=&\frac{3}{16\pi l^2}-\frac{1}{16\pi r_{+}^{2}}+\frac{3(q_{e}^{2}+q_{m}^{2})}{16\pi r_{+}^{4}}\nonumber\\
&+&\frac{3\alpha}{16\pi^{2} l^2 r_{+}^{6}}
(r_{+}^{4}-\frac{l^{2}}{3}r_{+}^{2}+(q_{e}^{2}+q_{m}^{2})l^{2})\ln{(\frac{4\sqrt{\pi}l^{2}r_{+}^{2}}{Ql^{2}-l^{2}r_{+}^{2}-3r_{+}^{4}})}.
\end{eqnarray}
The first term of r.h.s is corresponding to ordinary AdS black hole pressure given by the equation (\ref{s12}).\\
Then, we can calculate enthalpy using the equations (\ref{s13}), (\ref{s14}), (\ref{s16}) and the following relation,
\begin{equation}\label{s17}
H=E+PV=\frac{3r_{+}^{3}}{4l^{2}}+\frac{5}{12}r_{+}+\frac{3(q_{e}^{2}+q_{m}^{2})l^{2})}{4r_{+}}+H(\alpha),
\end{equation}
where $H(\alpha)$ is contribution of thermal fluctuations.
Enthalpy of the black hole may be considered as black hole mass \cite{11, 12}, from the left plot of the Fig. \ref{fig3} we can see as the $\alpha=0$ limit (dotted line) enthalpy behaves as black hole mass which illustrated by the Fig. \ref{fig1} (a). As expected we can see that enthalpy of large black hole has similar behavior for $\alpha=0$, $\alpha=1$, $q_{e}^{2}+q_{m}^{2}=1$ and $q_{e}^{2}+q_{m}^{2}=0$. For the uncharged black hole we can see that effect of thermal fluctuation is some negative enthalpy (negative mass) which is forbidden. Hence, as before, there is a minimum value for the horizon radius of the black hole. There is a critical value for the horizon radius $r_{c+}$, effect of thermal fluctuations is decreasing $H$ for $r_{+}>r_{c+}$ and is increasing $H$ for $r_{+}<r_{c+}$.\\

Then we can obtain Gibbs free energy using the following relation,
\begin{equation}\label{s18}
G=H-TS=\frac{r_{+}}{6}+\frac{Q}{r_{+}}+G(\alpha),
\end{equation}
where $G(\alpha)$ is contribution of thermal fluctuations. In the right plot of the Fig. \ref{fig3} we can see behavior of Gibbs free energy with variation of $\alpha$ and black hole charge. We can see similar behavior with Helmhotz free energy. Apparent kink in the right plot and initial value of left plot is about $r_{+c}=0.7$ as explained after the Fig. \ref{fig2}.\\
The modified first law of black hole thermodynamics of this case given by \cite{9},
\begin{equation}\label{first}
dM=Tds+\Phi_{e}dq_{e}+\Phi_{m}dq_{m}.
\end{equation}
It is clear that the case of $\alpha=0$ satisfied first law of black hole thermodynamics. It would be interesting to investigate validity of above equation in presence of logarithmic correction and obtain appropriate condition to satisfy the equation (\ref{first}). After a bit calculation we find that the following condition
\begin{equation}\label{condition}
q_{e}\frac{dq_{e}}{dr_{+}}+q_{m}\frac{dq_{m}}{dr_{+}}=(1+\frac{3r_{+}^{2}}{l^{2}})r_{+},
\end{equation}
is necessary to have valid first law of black hole thermodynamics.

\section{Holographic dual}
Dyonic charged AdS black hole in absence of thermal fluctuations is holographic dual of a fluid with van der Waals equation state given by \cite{mann},
\begin{equation}\label{s19}
\left ( P+ \frac{a}{V^2}\right )(V-b) = kT.
\end{equation}
where $k$ is the Boltzmann constant. The constant $ b > 0 $ denotes nonzero size
of the molecules of a given fluid, while the constant $ a > 0 $ is
a measure of the interaction between them. We will show numerically that the thermal fluctuations have not so important effect on P-V diagram, and discuss about critical points where,
\begin{eqnarray}\label{s20}
\left(\frac{\partial P}{\partial V}\right)_{T=T_{c}} &=& 0, \nonumber\\
\left(\frac{\partial^{2}P}{\partial V^2 }\right)_{T=T_{c}} &=& 0.
\end{eqnarray}
At $T=T_{c}$ direction of P-V line is zero but it is not minimum or maximum. Below it we have some region with negative compressibility corresponding to dual van der Waals fluid, which means instability of black hole.

\begin{figure}[h!]
 \begin{center}$
 \begin{array}{cccc}
\includegraphics[width=50 mm]{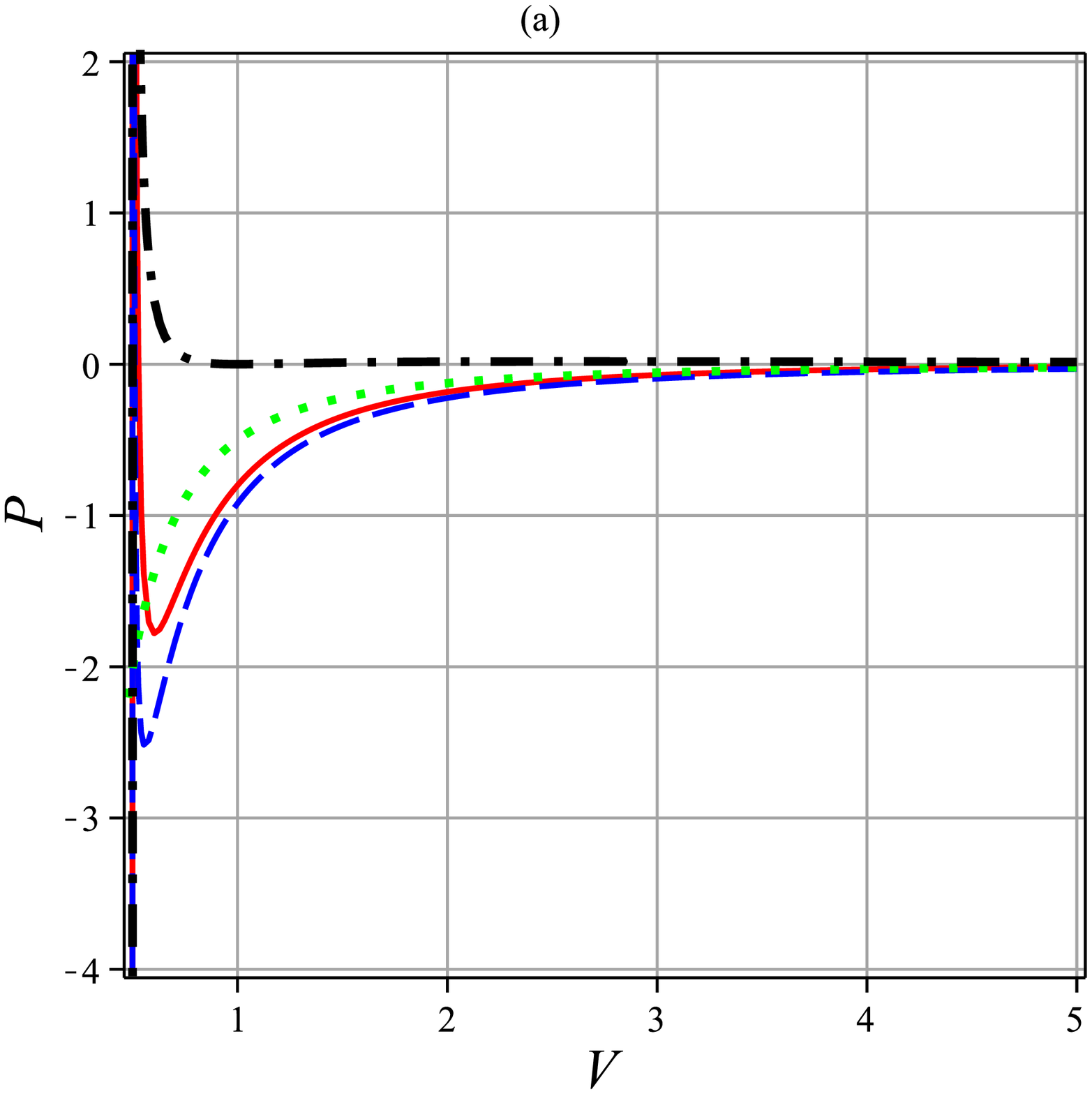}\includegraphics[width=50 mm]{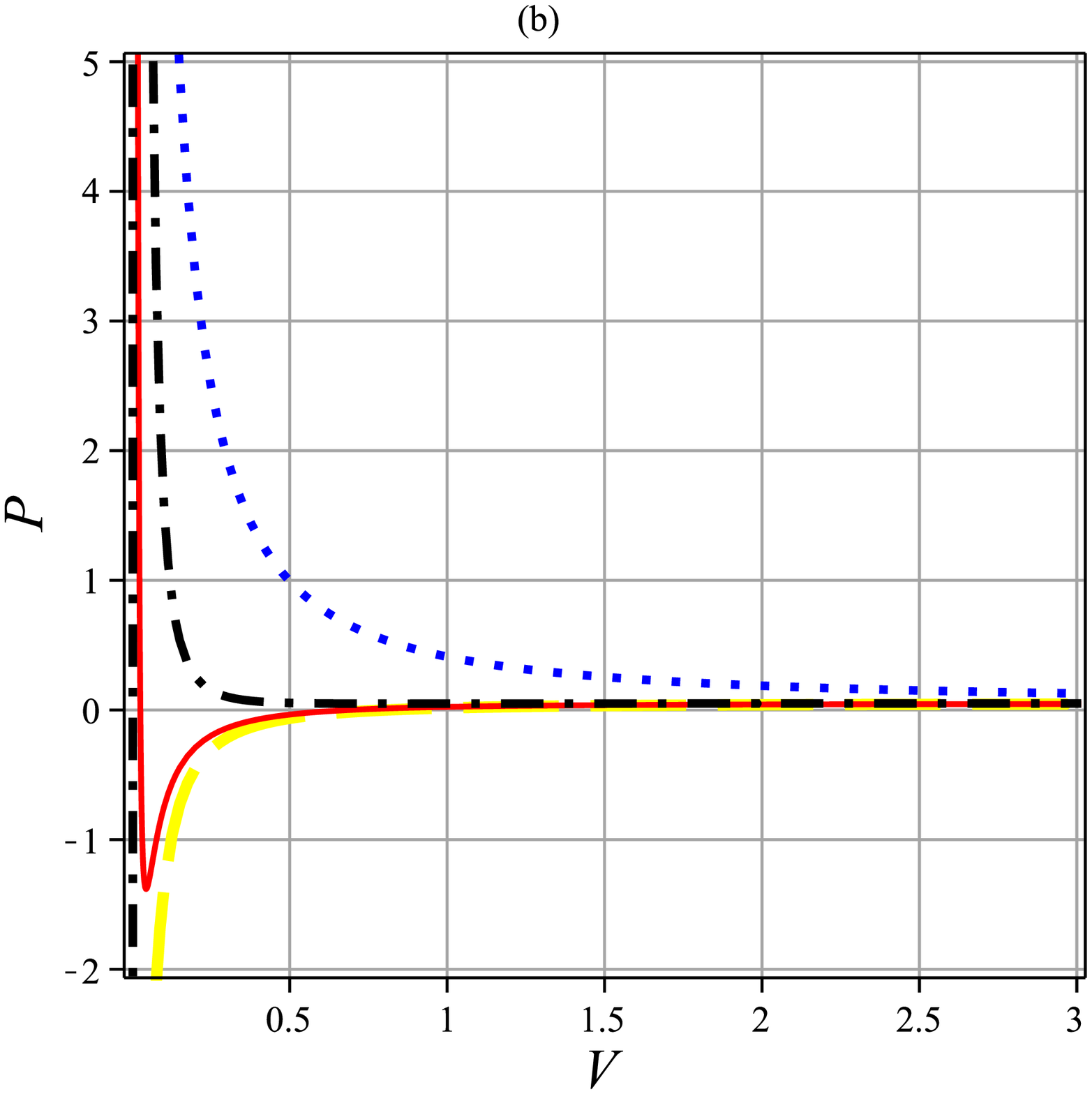}
 \end{array}$
 \end{center}
\caption{(a) P-V diagram of van der Waals fluid with $b=0.5$ and $k=0.1$. Dotted green: $T=0$, $a=1$. Dashed blue: $T=0.4$, $a=1$. Solid red: $T=1$, $a=1$. Dash dotted black: $T=1$, $a=0.2$. (b) P-V diagram of dyonic charged AdS black hole with $l=1$. $Q\equiv q_{e}^{2}+q_{m}^{2}=0.04$, $\alpha=1$ (dash dotted black), $\alpha=0$ (dotted blue). $Q=0$, $\alpha=1$ (dashed yellow).  $Q\equiv q_{e}^{2}+q_{m}^{2}=0.01$, $\alpha=1$ (solid red).}
 \label{fig4}
\end{figure}

The typical behavior of P-V diagram corresponding to van der Waals fluid plotted in Fig. \ref{fig4} (a).\\
We can use equation (\ref{s13}) and (\ref{s16}) to investigate P-V diagram of dyonic charged AdS black hole in presence of thermal fluctuations. Our numerical analysis illustrated by Fig. \ref{fig4} (b). We can see solid red lines in both (a) and (b) plots of Fig. \ref{fig4} to find that dyonic charged AdS black hole in presence of thermal fluctuations with suitable value of electric and magnetic charges is also dual of van der Waals fluid.  We find that value of charge is many important to have dual van der Waals fluid. Therefore, thermodynamics quantities obtained in the previous section use to see thermal fluctuation and hence quantum gravity effects.\\
We will discuss about critical points (corresponding to black lines (dash dot) of both Fig. \ref{fig4} (a) and (b)) and stability of the model in the next section.\\

\section{Critical points and stability}
\begin{figure}[h!]
 \begin{center}$
 \begin{array}{cccc}
\includegraphics[width=50 mm]{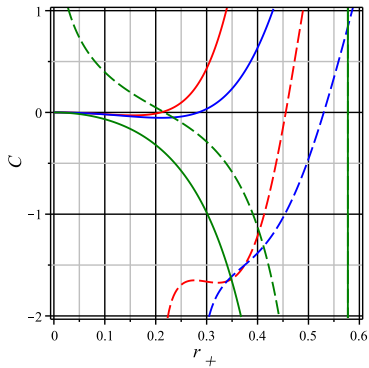}
 \end{array}$
 \end{center}
\caption{Specific heat $C$ in terms of horizon radius with $l=1$. $Q\equiv q_{e}^{2}+q_{m}^{2}=0.1$, $\alpha=1$ (dashed blue), $\alpha=0$ (solid blue). $Q=0.05$, $\alpha=1$ (dashed red), $\alpha=0$ (solid red). $Q=0$, $\alpha=1$ (dashed green), $\alpha=0$ (solid green).}
 \label{fig5}
\end{figure}

Critical points of the model are corresponding to the dash dotted black lines of Fig. \ref{fig4} (a) and (b). At $\alpha=0$ limit and using both conditions given by (\ref{s20}) one can obtain the following equation,
\begin{equation}\label{s20-1}
r^{3}+3r^{2}-6(q_{e}^{2}+q_{m}^{2})r-30(q_{e}^{2}+q_{m}^{2})=0.
\end{equation}
It gives us critical horizon radius as follow,
\begin{equation}\label{s20-2}
r_{+c}(\alpha=0)=\frac{f(Q)^{2}-f(Q)+2(q_{e}^{2}+q_{m}^{2})+1}{f(Q)},
\end{equation}
where we defined,
\begin{equation}\label{s20-3}
f(Q)=\left[12(q_{e}^{2}+q_{m}^{2})-1+\sqrt{132(q_{e}^{2}+q_{m}^{2})^{2}-8(q_{e}^{2}+q_{m}^{2})^{3}-30(q_{e}^{2}+q_{m}^{2})}\right]^{\frac{1}{3}},
\end{equation}
which means that $0.23\leq(q_{e}^{2}+q_{m}^{2})\leq16.27$. However we expect that effect of thermal fluctuations change charge interval.\\
Specific heat is an important measurable physical quantity which can determine stability of system. It has the following relation with internal energy $E$, entropy $S$ and temperature $T$,
\begin{equation}\label{s21}
C= T \left (\frac{\partial S}{\partial T} \right )=\frac{\partial E}{\partial T}.
\end{equation}
If $C>0$ black hole is stable and if $C<0$ black hole is in unstable phase. So, $C=0$ is corresponding to phase transition of van der Waals fluid, similar to the critical point discussed above. By using the equation (\ref{s14}) we can obtain specific heat as following,
\begin{equation}\label{s22}
C= \frac{2(3r_{+}^{4}+l^{2}r_{+}^{2}-(q_{e}^{2}+q_{m}^{2})l^{2})\pi r_{+}^{2}}{3(r_{+}^{4}-\frac{1}{3}l^{2}r_{+}^{2}+(q_{e}^{2}+q_{m}^{2})l^{2})}+C(\alpha),
\end{equation}
where $C(\alpha)$ is logarithmic corrected terms. In the Fig. \ref{fig5} we can see effect of thermal fluctuations on the stability of black hole. We can see that there is a minimum size for the black hole as expected. Below minimum size $r_{+}<r_{+c}$, black hole in unstable and it means that van der Waals fluid changes phase. Fortunately, it is possible to see thermal fluctuation effects for $r_{+}\geq r_{+c}$ where black hole is stable and we have van der Waals fluid. Reducing electric and magnetic charges help to obtain smaller minimum size, therefore we have more chance to see effects of thermal fluctuations. For the case of $Q=1$ one may have $r_{+c}\approx0.7$ as discussed after the Fig. \ref{fig2} and Fig. \ref{fig3}. Without thermal fluctuations it is clear that the black hole can exist at smaller size which illustrated by the Fig. \ref{fig5} (compare solid and dashed red lines for example). We see that while thermal fluctuations reduce stability of black hole but it has enough stable region to reflect quantum gravity effects. Also, green lines represent uncharged case which is completely unstable without logarithmic correction and we can see opposite behavior comparing with charged black hole.

\section{Conclusion}
In present work we considered special solutions of Einstein-Maxwell theory with negative cosmological constant in four dimensions which is dyonic charged AdS black hole. Our main work is finding the effect of thermal fluctuations on the thermodynamics quantities.\\
Thermal fluctuations exist for any black object, but they are important for small black hole and negligible for the large black hole. Advantage of dyonic charged AdS black hole background is its holographic picture which is a van der Waals fluid. We have shown that, in presence of thermal fluctuations there is still a van der Waals fluid as dual picture. Only we should fix black hole charge which is corresponding to the electric charges of a van der Waals fluid. We obtained some thermodynamics quantities like Gibbs and Helmholts free energies and shown that thermal fluctuations have not important effects for the large black hole. On the other hand for the small black holes they are important and have crucial role. We found that thermal fluctuations reduced stable regions of the black hole, however there are enough stable regions to see quantum gravity effects before phase transition of a van der Waals fluid. It means that there is a minimum radius, we called critical radius, where black hole is stable in presence of thermal fluctuations, and in this region dyonic charged AdS black hole in presence of thermal fluctuations is dual of van der Waals fluid.\\\\\\

{\bf Acknowledgments} It is pleasure to thanks Prof. Robert B. Mann for introducing origin of holographic dual of charged AdS black hole. Also we would like to thanks S. Mahapatra and Yu Tian for useful comments.

\end{document}